# Linear and Nonlinear Optical Propagation in 2D Materials

Nicola Curreli*,[1] Alessandro Fanti,[2] Giuseppe Mazzarella,[2] and Ilka Kriegel[1]

[1] *Functional Nanosystems, Istituto Italiano di Tecnologia, via Morego 30, 16163 Genova, Italy*
[2] *Department of Electric and Electronic Engineering, University of Cagliari, Via Marengo 2, 09123 Cagliari, Italy*

Recently, a lot of effort has been dedicated to developing next-generation optoelectronic devices based on two-dimensional materials, thanks to their unique optical properties that are significantly different from those of their bulk counterparts. In order to implement high-performance nanoscale optical devices, an in-depth study of how linear and non-linear propagation occurs in two-dimensional materials is required. Here, we focus on the theory behind the propagation of electromagnetic waves in two-dimensional materials as well as emerging applications in the fields of electronics, optics, sensors, which are summarized and discussed in the paper.

## INTRODUCTION

Over the last 15 years, the study of two-dimensional (2D) materials has progressed rapidly in a variety of scientific and engineering subfields [1, 2]. Compared to their conventional bulk counterparts, 2D materials exhibit electronic confinement effects within two dimensions because their thickness is in the order of subnanometers [3–5]. The types of 2D materials available have been continuously growing, including insulators, semiconductors, and a plethora of metals and semimetals [6–22]. Their tunable optical and electronic properties [23–25] and their easy integration allow improving the performance of optoelectronic devices [26–30]. In fact, 2D materials properties are determined by their structural features, in which dimensionality plays a fundamental role [31, 32]. When approaching the two-dimensional nature of crystals, defined as an infinite crystalline in-plane periodic structure with atomic thickness, novel peculiar properties arise [33]. In addition to unique electronic and optical properties, these materials show high mechanical strength and flexibility [34]. The quantum confinement in the direction perpendicular to 2D crystals planes, promote a longer mean free path of electrons, excitons, phonons, and ballistic in-plane transport without scattering or diffusion [31, 35, 36]. Moreover, the 2D geometry is compatible with the design and implementation of optical devices finding application in electronics, optics, and many other fields [37]. For these reasons, in this article, we focus on theoretical and experimental research of linear and non-linear optical propagation and effects caused by different electromagnetic external stimuli in 2D materials.

## ELECTRONIC AND OPTICAL PROPERTIES OF 2D MATERIALS

Since the discovery of graphene [38], new materials have been added to the family of 2D materials at an increasing pace. Among them, the most interesting are hexagonal boron nitride (h−BN) [31, 39, 40], transition-metal dichalcogenides (TMDs, such as $MoS_2$, $MoSe_2$, $WS_2$, $WSe_2$) [40, 41], silicene [42, 43], germanene [44], stanene [45], phosphorene [46–48], borophene [49, 50], just to cite few. All of those materials cover the entire spectrum of electrical conductivity from (semi)metals (*e.g.*, graphene and borophene) and semiconductors (*e.g.*, TMDs) to insulators (*e.g.*, h−BN) (Fig. 1).

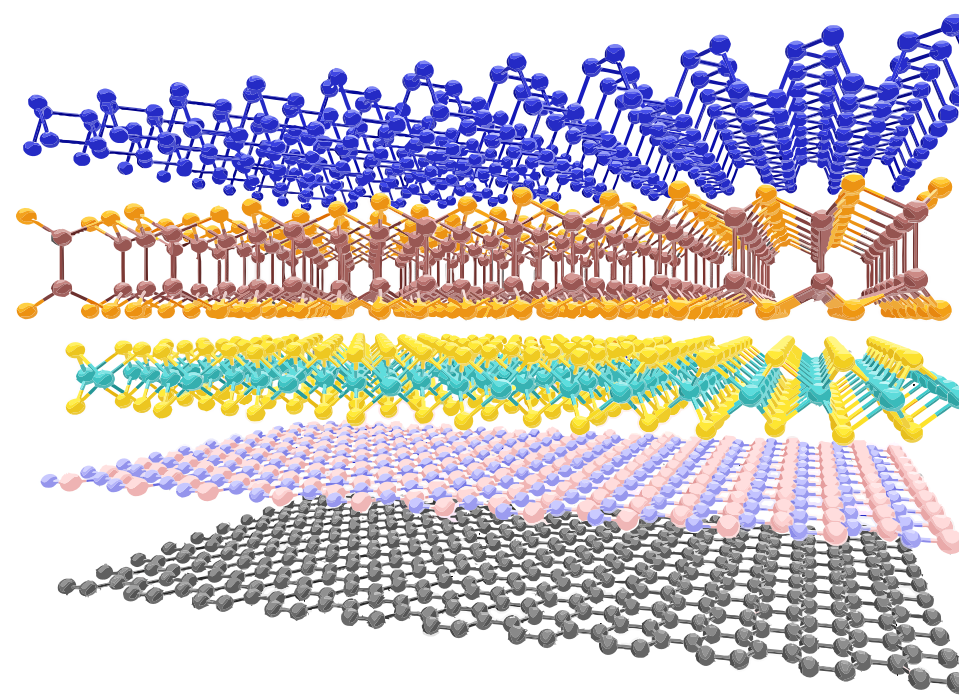

**FIG. 1:** Sketch of several 2D materials. From top to bottom: phosphorene, InSe, $MoS_2$, h−BN and graphene.

### Graphene

Graphene is a 2D crystal formed by carbon atoms arranged in a hexagonal honeycomb lattice. The electronic structure of graphene can be derived through a tight-binding model, in which Dirac cones are found at the corners of the Brillouin zone (Fig. 2a) [51, 52]. In the proximity of these points, charge carriers obey a linear dispersion relation:

$$E = \pm v_F p \tag{1}$$

where $E$ and $p$ are the energy and the momentum measured with respect to the Dirac points, $v_F = 10^6\ ms^{-1}$ is the Fermi velocity, while the plus and minus signs refer to the conduction and valence bands, respectively [52]. The bands with conical dispersion intersect at the Fermi energy, thus making graphene a semimetal.

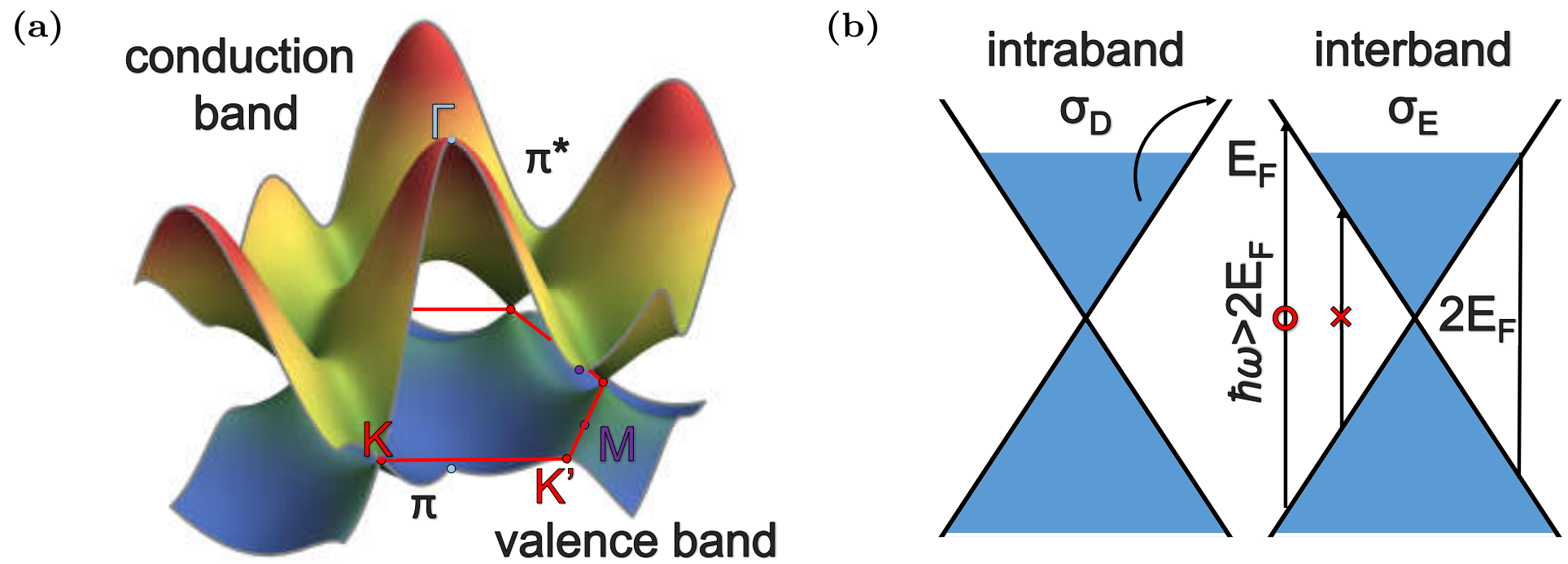

**FIG. 2:** a) The electronic energy dispersion of graphene throughout the whole region of Brillouin zone. b) The dispersion around K point, on the left the intraband transition of electron while, on the right the interband transition of electron, which occurs if $\hbar\omega > 2E_F$.

As a result, an additional-chiral-symmetry exists, fixing a parallel or antiparallel pseudospin to the directions of motion of the carriers [52, 53]. This contributes to peculiar effects on the electronic and optical properties of graphene. For example, for undoped samples at $0\ K$ the universal optical conductivity of graphene, which links the surface current density $\boldsymbol{J}$ to the electric field $\boldsymbol{E}$ ($\boldsymbol{J} = \bar{\sigma}\boldsymbol{E}$), is independent of any material parameter:

$$\bar{\sigma}_0 = \frac{e^2}{4\hbar} \sim 6.08 \times 10^{-5}\ S \tag{2}$$

where $e$ is the electron charge and $\hbar$ is the reduced Planck's constant. Thus, the transmission for a single layer graphene is given as [53–56]:

$$T = (1 + 0.5\pi\alpha)^{-2} \sim 1 - \pi\alpha \sim 97.7\% \tag{3}$$

where $\alpha = e^2/(4\pi\varepsilon_0\hbar c) = \bar{\sigma}_0/(\pi\varepsilon_0 c) \sim 1/137$ is the fine-structure constant [54, 56]. Notably, the optical absorption depends only on the fine-structure constant, and is given as $A \sim 1 - T \sim \pi\alpha \sim 2.3\%$, considering that for a monolayer the reflection contribution ($R$) is less than 0.1% (Fig. 3) [56]. On the other hand, doping has a strong effect on optical properties [53, 57], because of the so-called Pauli blocking [53]. In fact, Pauli blocking ensures that photons with energy ($\hbar\omega$) less than $2E_F$ (where $E_F$ is the Fermi energy) can not be absorbed (Fig. 2b) [52–54, 58].

The conical dispersion of low-energy carriers in graphene is very different from other materials' parabolic dispersion [51, 52]. For example, bilayer graphene presents parabolic dispersion at the Fermi energy and a small bandgap of up to 250 $meV$ that can be opened with an electric field [54, 59, 60]. In the case of multiple layer structures (such as bilayer or few-layer graphene), it is possible to consider an optically equivalent superposition of almost non-interacting single-layer graphene flakes in which each layer can be seen as a 2D electron gas with little perturbation from the adjacent one [61]. For few-layer graphene, the contribution of the reflection becomes more prominent, reflecting $\sim 2\%$ of the incident light for ten-layer crystals [61].

Frequency-independent absorption of graphene along with extremely high carrier mobility has attracted the interest of the scientific community. A particularly interesting research direction is its use in flexible electronic devices such as touch screens [54], electronic paper, organic photovoltaic cells [62] and organic light-emitting diodes [63], thanks to its flexibility as well as its low surface resistance and high transmittance. Moreover, graphene is used to implement photodetectors that are used in the broadband spectral region between the ultraviolet and infrared [64]. Furthermore, in a single layer of graphene [65], the interband transition photoelectrons are modulated by a driving voltage over a wide band in order to obtain an optical modulator with a bandwidth in the near-infrared region of over 1 GHz [64].

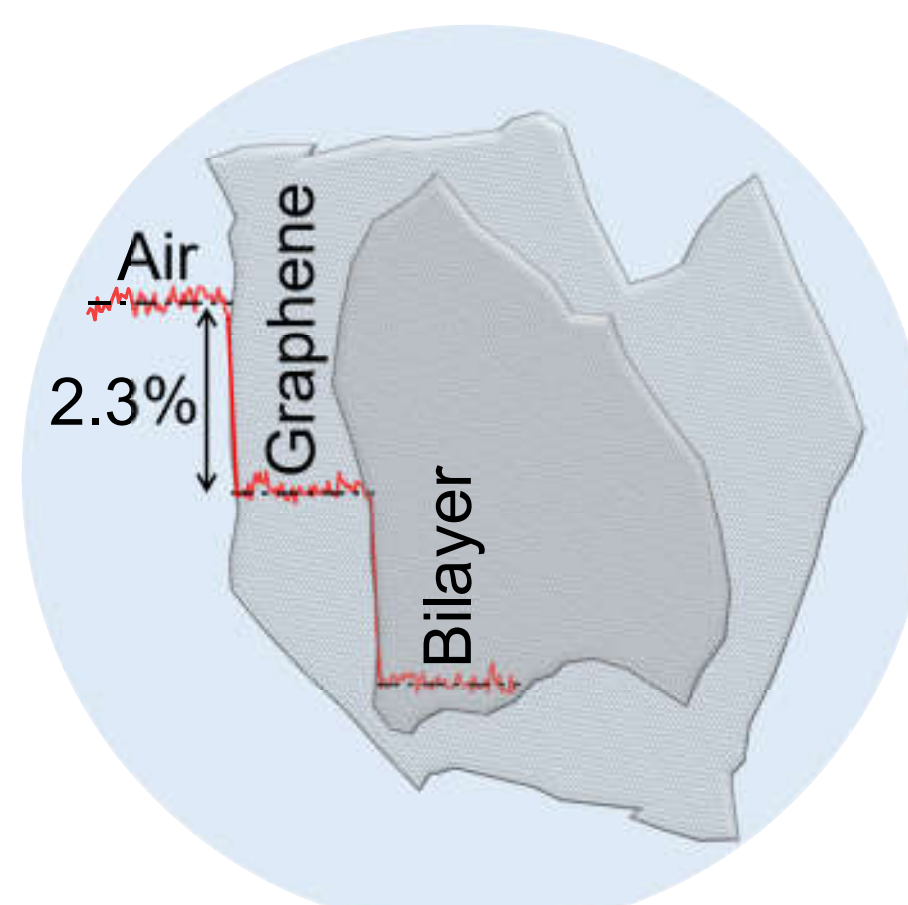

**FIG. 3:** Schematic of the intensity of transmitted white light through graphene and bilayer graphene. The dashed line scan profile shows the intensity between layers.

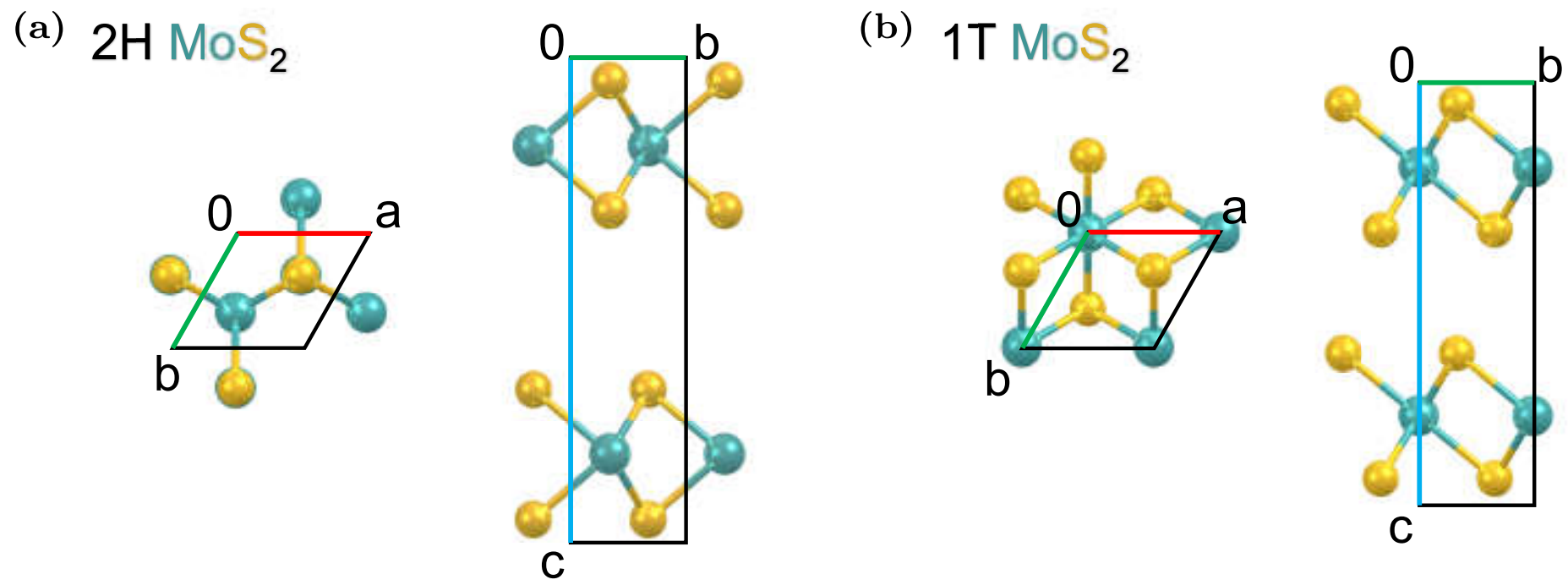

**FIG. 4:** a) 2H and b) 1T $MoS_2$ crystal structure.

### Transition Metal Dichalchogenide Monolayers

The interesting properties of graphene have led to the discovery and study of other 2D crystals and heterostructures thereof, showing exciting optical properties which differ from their bulk counterparts. Transition metal dichalcogenides (TMDs) are probably the second most-studied 2D material class after graphene. They are a family of materials with $MX_2$ chemical formula, where M is a transition metal and X is a chalcogen such as S, Se, or Te [41, 51]. The TMDs' atomic and electronic structure consists of a layer of M atoms inserted between two atomic layers of X atoms, with a hexagonal unit cell having $D_{3h}$ symmetry and a thickness of $\sim 0.7\ nm$ [41, 66–68]. Within the TMDs it is possible to find metallic (*e.g.*, $VS_2$ and $NbS_2$), semiconducting (*e.g.*, $MoS_2$ and $WS_2$) or insulating (*e.g.*, $HfS_2$) materials [41, 68]. The TMD crystals show different polyforms [41, 68]. Typical crystal structures include trigonal prismatic 1H (2H for multilayer), 1T, and 1T' (*i.e.*, distorted-1T) phases [33, 37, 69], as shown in Fig. 4. The most stable crystalline configuration is the 2H-stacking (also known as AB-stacking) [41, 51, 70]. In particular, semiconducting 2H TMDs show peculiar optical and electronic properties, fundamentally different from their 3D bulk counterparts. For example, $MoS_2$ [66, 67, 71], $MoSe_2$ [51, 72], $WS_2$, and $WSe_2$ [73] undergo a crossover from indirect to direct gap when going from bilayer to monolayer form. In the case of $MoS_2$, bulk crystals have an indirect bandgap of the order of 1.29 eV, while in its 2D form, $MoS_2$ presents a direct bandgap [53, 71] of $\sim 1.9\ eV$ around the K points of the Brillouin zone. This leads to pronounced luminescence enhancement [66].

### Other 2D materials

Similarly, layered post-transition metal chalcogenides of the group-III, such as GaS, GaSe, SnS, $SnS_2$, and InSe, are also being explored due to their high mobilities, large photoresponses, and in-plane anisotropy [70, 74]. They present different polytypes which vary with respect to their layer-stacking configuration, namely, the group-IV monochalcogenides (*e.g.*, SnS and GeSe) [75] or the 1T dichalcogenides (*e.g.*, $SnS_2$) [70].

Moreover, there is also the family of 2D transition metal carbides, nitrides, and carbonitrides which are known as MXenes with the following chemical formulas: $M_2X$, $M_3X_2$ and $M_4X_3$, where M is an early transition metal and X is a carbon or a nitrogen atom [76]. They exhibit interesting mechanical properties as well as good thermal and electrical conductivity [70].

In addition, wide bandgap 2D materials play a key role in device implementation. For example, hexagonal boron nitride (h−BN), thanks to its inert nature, insulating characteristics and ultra-flat structure, act as a substrate or sacrificial layer for high-mobility 2D crystal-based devices [77].

Other wide bandgap materials, including transition metal oxides (TMO - *e.g.*, $MoO_2$ with 2H-phase, $MnO_2$ with 1T-phase and $\alpha$-$MoO_3$) and chromium oxide (*i.e.*, $Cr_2O_3$), known for its multiferroic properties [78], exhibit hyperbolic optical behavior [70, 79].

Other examples of layered materials include elementary 2D materials, such as phosphorus allotropes (*e.g.*, black and blue phosphorus) [47], silicene [80], germanene [81], tellurene [82], gallenene [83], antimonene [84], and borophene [85], which range from metallic to semiconductor as well as topological insulators (for example, $Bi_2Se_3$ and $Sb_2Se_3$) known for their topologically protected and spin-momentum-locked electronic transport [70, 86].

## OPTICAL PROPAGATION IN 2D MATERIALS

In this section, the electromagnetic surface wave [87] that propagates in a direction parallel to the 2D material layer, as well as transmission and reflection of electromagnetic waves by a periodic structure consisting of N layers, will be investigated.

### The electromagnetic surface wave in 2D materials

An electromagnetic surface wave is a field configuration essentially confined in a small region close to the interface between two dielectric media and propagating in the direction along the interface itself [88–94]. Here, we consider only plane interfaces and choose a cartesian reference frame with the $z$-axis normal to the interface (Fig. 5). By our convention, the surface wave propagates in the $x$-direction and the field decays in the $z$-direction. The basic principle behind the phenomenon of the surface waves is the solution of Maxwell's equations at the interface. From this, it is possible to obtain that surface waves are excited by external incident electromagnetic waves [88, 94, 95]. Surface plasmon polaritons (SPPs) are a particular type of electromagnetic surface waves traveling along with a metal-dielectric or metal-air interface in the infrared or visible-frequency range. The physical characteristics of SPPs in metals are determined by their intrinsic properties and the geometry of the material. The thickness plays a very important role in the physical propagation of SPPs. In fact, more recently, it has been found that 2D materials, inserted at the interface between dielectrics, support SPPs thanks to the strongly confined electric field as the thickness approaches the atomic level and light is being confined to dimensions 2-3 order of magnitude smaller than that of the free-space wavelength [96]. The polarization of the incident waves determines the polarization of surface waves, therefore it is possible to find transverse magnetic (TM) and transverse electric (TE) surface waves, determining the components of the field for each surface wave. In the TM surface wave, the incident field has a component of a magnetic field perpendicular to the incidence plane (Fig. 5a), while in the TE surface wave, the incident field has a component of an electric field perpendicular to the incidence plane (Fig. 5b).

#### *Field configuration at TM surface wave*

For a TM surface wave, $E_x$, $E_z$, $H_y$ field components are considered, and it is assumed that the 2D material is surrounded by two dielectric media 1 and 2 as shown in Fig. 5a. The TM surface wave propagates on the surface of the 2D material in the $x$-direction with wave vector in the direction of propagation ($\overline{q} = q\overline{x}$). In medium 1, the electromagnetic fields of the TM surface wave are given by [88]:

$$\begin{aligned} E_x^{(1)} &= E_1 e^{iqx} e^{-\kappa_1 z} \\ E_z^{(1)} &= \frac{iq}{\kappa_1} E_x^{(1)} \\ H_y^{(1)} &= -\frac{i\omega\varepsilon_0\varepsilon_1}{\kappa_1} E_x^{(1)} \end{aligned} \tag{4}$$

while in medium 2:

$$\begin{aligned} E_x^{(2)} &= E_2 e^{iqx} e^{\kappa_2 z} \\ E_z^{(2)} &= -\frac{iq}{\kappa_2} E_x^{(2)} \\ H_y^{(2)} &= \frac{i\omega\varepsilon_0\varepsilon_1}{\kappa_2} E_x^{(2)} \end{aligned} \tag{5}$$

where $E_x^{(i)}$, $E_z^{(i)}$ and $H_y^{(i)}$ are the electric field in the $x$, $y$ and $z$ directions respectively, $E_i$ is the amplitude of electric field in the $x$-direction, for the $i^{th}$ medium ($i =$ 1, 2), $\varepsilon_0$ and $\varepsilon_1$ are the dielectric constant of the vacuum and the medium respectively, and $\kappa_i$ is decay constant of the fields inside the $i^{th}$ medium given by:

$$\kappa_i = \sqrt{q^2 - \omega^2\varepsilon_i/c^2} \tag{6}$$

The magnetic and the electric field are related by the following relations:

$$E_x^{(i)} = -\frac{i}{\omega\varepsilon_0\varepsilon_i}\frac{\partial H_y^{(i)}}{\partial z} \tag{7}$$

$$E_z^{(i)} = -q/(\omega\varepsilon_0\varepsilon_i) H_y^{(i)} \tag{8}$$

which can be derived from Maxwell's equations [88]. Due to the presence of surface currents that flow on the 2D material, the magnetic field components are not continuous. Therefore, the boundary conditions of the electromagnetic wave on the surface of 2D material are given by:

$$E_x^{(1)} = E_x^{(2)} \tag{9}$$

$$H_y^{(1)} - H_y^{(2)} = -J \tag{10}$$

where $J = \bar{\sigma}(\omega) E_x^{(2)}$ is the surface current on the 2D material and $\bar{\sigma}(\omega)$ is the complex optical conductivity of 2D material. Since a 2D material presents a negligible thickness, when compared to the wavelength in the materials, $J$ appears only in the boundary conditions.

In the case of graphene, there are two contributions to $\bar{\sigma}(\omega)$, corresponding to two possible optical scattering or excitation of electron by photon, which are the intraband transition (within the same conduction band) and interband transition (from valence to conduction band), as shown in Fig. 2b. The real and imaginary parts of $\bar{\sigma}(\omega)$ are related by the Kramers-Kronig relation and it is given by the following equation [97–102]:

$$\begin{aligned} \bar{\sigma}(\omega) &\equiv \sigma_D + \mathrm{Re}\{\sigma_E\} + i\,\mathrm{Im}\{\sigma_E\} = \\ &= \frac{E_F e^2}{\pi\hbar}\frac{i}{\hbar\omega + i\Gamma} + \frac{e^2}{4\hbar}\delta_{-1}(\hbar\omega - 2E_F) + \\ &\quad - \frac{ie^2}{4\pi\hbar}\ln\left|\frac{\hbar\omega + 2E_F}{\hbar\omega - 2E_F}\right| \end{aligned} \tag{11}$$

The first term in Eq. 11 ($\sigma_D$) is the intraband conductivity (or Drude conductivity) [99–101], while the second and the third terms correspond to the real part and the imaginary part of interband conductivity ($\sigma_E$), respectively [97, 99, 101]. The $\delta_{-1}(x)$ is the Heaviside step function, while the spectral width ($\Gamma$) is a parameter that describes the scattering rate with phonons due to the impurities and depends on the Fermi level ($E_F$) as:

$$\Gamma = \hbar e v_F^2/\mu E_F \tag{12}$$

where $v_F = 10^6\ m/s$ is the Fermi velocity of graphene and $\mu = 10^4\ cm^2/Vs$ is the mobility for ideal graphene [100]. For $\hbar\omega < 2E_F$ the intraband transition is dominant and the optical transition of the electron with $q \neq 0$ is possible, due to the additional scattering of an electron by impurity or phonon, which might modify the momentum of the electron. On the other hand, the interband transition from the valence to conduction band is dominant for $\hbar\omega > 2E_F$ [101]. For a 2D electron gas system (*e.g.*, GaAs/AlGaAs quantum-well structure), the optical conductivity is described by [97, 103]:

$$\bar{\sigma}_{2Dgas} = i\frac{ne^2}{m(\omega + i\gamma)} \tag{13}$$

where $n$, $m$ and $\gamma$ are the density, effective mass, and scattering rate of the 2D electron gas system. The dispersion of TM surface wave is obtained by substituting Eqs. 4-5 to boundary conditions (Eqs. 9 and 10). In addition, the following equation is a requirement to have TM surface wave in the 2D material [97, 99, 100, 104]:

$$\frac{\varepsilon_1}{\kappa_1} + \frac{\varepsilon_2}{\kappa_2} + \frac{i\bar{\sigma}(\omega)}{\omega\varepsilon_0} = 0 \tag{14}$$

where $\varepsilon_i$, ($i$ = 1,2) is the dielectric constant of the $i^{th}$ medium, which surrounds the 2D material. Considering that $\varepsilon_i > 0$ and $\kappa_i > 0$. In order to satisfy Eq. 14, the imaginary part of $\bar{\sigma}(\omega)$ should be positive ($\text{Im}\{\bar{\sigma}(\omega)\} > 0$). This requirement is satisfied by the conventional 2D electron gas system, since $\text{Im}\{\bar{\sigma}_{2Dgas}\} = ne^2\omega/(m(\gamma^2 + \omega^2)) > 0$ [97], and in the case of graphene $\text{Im}\{\bar{\sigma}(\omega)\} > 0$ for $\hbar\omega < 1.667E_F$. By solving Eq. 14, it is possible to derive the dispersion relation of TM surface wave for graphene. If $\hbar\omega << 2E_F$ (being $E_F = 0.64\ eV$), it is obtained that $\bar{\sigma}(\omega) \sim \sigma_D(\omega)$ and $\Gamma$ can be ignored, since it gives only the spectral broadening of the TM surface wave. When the velocity of light can be considered to be much larger than the group velocity of surface plasmon $v_{sp}$, thus $q >> \omega/c$ obtaining $\kappa_1 = \kappa_2 = q$ in Eq. 6 this condition is called the non-retarded regime [93, 100, 105]. By substituting $\sigma_D$ in Eqs. 11 to Eq. 14 and solve for $\omega$, a square-root dependence ($\omega \propto \sqrt{q}$) of the dispersion relation is obtained [93, 100, 105, 106]:

$$\omega = \frac{1}{\hbar}\sqrt{\frac{E_F e^2 q}{\pi\varepsilon_0(\varepsilon_1 + \varepsilon_2)}} \tag{15}$$

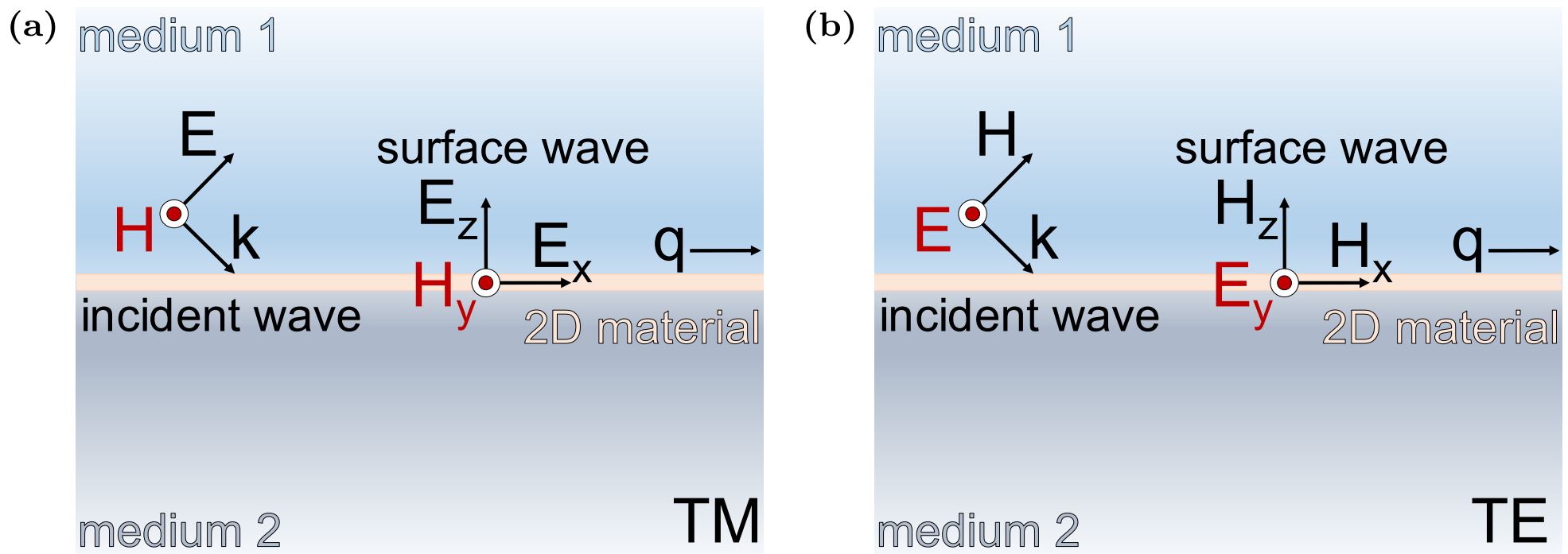

**FIG. 5:** a) TM surface wave and b) TE surface wave in 2D materials. 2D material is shown as orange color with negligible thickness.

The aforementioned relation can be also obtained for conventional 2D electron gas by substituting Eq. 13 to Eq. 14, solving for $\omega$, finding that $\omega \propto \sqrt{q}$ dependence is a characteristic of SPP in 2D materials [93, 98, 100, 105]. Considering graphene as a medium, the long-wavelength plasmon dispersion can be calculated by looking for zeros of the dynamical dielectric function ($\varepsilon_g(q,\omega) = 0$) [98, 105].

*Field configuration at TE surface wave*

For a TE surface wave (Fig. 5b), the field components are $E_y$, $H_x$, $H_z$. In medium 1, the electromagnetic fields of the TE surface wave are given by:

$$\begin{aligned} H_x^{(1)} &= H_1 e^{iqx} e^{-\kappa_1 z} \\ H_z^{(1)} &= \frac{iq}{\kappa_1} H_x^{(1)} \\ E_y^{(1)} &= \frac{i\omega\mu_0}{\kappa_1} H_x^{(1)} \end{aligned} \tag{16}$$

and in medium 2:

$$H_x^{(2)} = H_2 e^{iqx} e^{\kappa_2 z}$$
$$H_z^{(2)} = -\frac{iq}{\kappa_2} H_x^{(2)}$$
$$E_y^{(2)} = -\frac{i\omega\mu_0}{\kappa_2} H_x^{(2)} \quad (17)$$

where $\mu_0$ is the permeability of the vacuum. Similarly for the TM surface wave, the magnetic and the electric field are related by the following relations:

$$H_x^{(i)} = \frac{i}{\omega\mu_0} \frac{\partial E_y^{(i)}}{\partial z} \quad (18)$$
$$H_z^{(i)} = q/(\omega\mu_0) E_y^{(i)} \quad (19)$$

and, by using the boundary conditions of the electromagnetic field at the surface $z = 0$, which are related to the case of TM surface wave given in Eqs. 9 and 10:

$$E_y^{(1)} = E_y^{(2)} \quad (20)$$
$$H_x^{(1)} - H_x^{(2)} = -J \quad (21)$$

at $z = 0$, where $J = \bar{\sigma}(\omega) E_y^{(2)}$ and assuming that the two dielectric media as vacuum ($\varepsilon_1 = \varepsilon_2 = 1$, thus $\kappa_1 = \kappa_2 = \sqrt{q^2 - (\omega/c)^2} \equiv \kappa$, it is obtained the following equation [97, 107]:

$$2 - \frac{i\bar{\sigma}(\omega)\omega\mu_0}{\kappa} = 0 \quad (22)$$

The concepts of surface plasmon dispersion and the non-retarded regime [100, 105] have been introduced in Eq. 14 and Eq. 15. For small $q$, the non-retarded approximation cannot be used because the group velocity of surface plasmon ($v_{sp}$) is comparable to the velocity of light. This case is the so-called retarded regime [105], in which the dispersion of surface plasmon is linear ($\omega \propto q$). From Eq. 6 and solving Eq. 14, supposing that $\varepsilon_1 = \varepsilon_2 = \varepsilon$, for retarded regime it has been obtained:

$$\omega \sim \frac{c}{\sqrt{\varepsilon}} q \quad (23)$$

For a conventional 2D electron gas system, $\mathrm{Im}\{\bar{\sigma}_{2Dgas}\} > 0$, as given by Eq. 13 [97], hence, the TE surface wave cannot be supported by a normal bulk metal or any other common material found in nature, since $\omega > 0$, and Eq. 23 requires $\mathrm{Im}\{\bar{\sigma}\} < 0$ [89, 97]. However, it is possible to obtain artificial materials that possess a negative permeability (*e.g.*, metamaterials [108] - generally not easy to fabricate - and 2D materials [109]). For example, in the case of graphene, $\mathrm{Im}\{\bar{\sigma}(\omega)\} < 0$ for $\hbar\omega < 2E_F$. This unusual property has enabled graphene to have the TE surface wave, although TE surface wave in doped graphene occurs only for a narrow frequency range of $1.667E_F < \hbar\omega < 2E_F$ [89, 97, 99, 110].

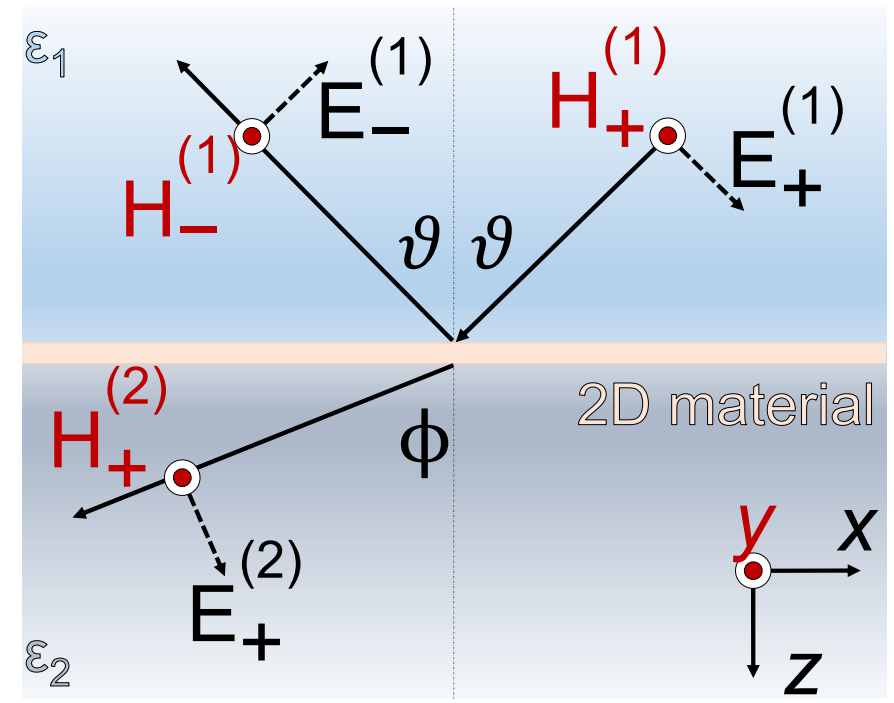

**FIG. 6:** A 2D material is placed between two dielectric media with dielectric constants $\varepsilon_1$ and $\varepsilon_2$. The thickness of the 2D material is neglected. The incident EM wave comes at an angle $\vartheta$ in medium 1 (top) and is transmitted at an angle $\phi$ in medium 2 (bottom).

### The optical absorption, reflection, and transmission spectra in 2D materials

To describe the optical spectra of the electromagnetic wave incident on a 2D material, Maxwell's equations with boundary conditions were first evaluated, considering the 2D material between two dielectric media. Subsequently, using the transfer matrix method, optical spectra were obtained in the case of a 2D material within a multilayer of dielectric media.

#### *2D material between two dielectric media*

The probabilities for absorption, reflection, and transmission of an electromagnetic wave penetrating 2D materials can be obtained by solving Maxwell's equations for an electromagnetic wave with boundary conditions. A 2D material, having a negligible thickness, is placed between two dielectric media as shown in Fig. 6. It is modeled as a conducting interface with a conductivity $\bar{\sigma}$ between two dielectric media with dielectric constants $\varepsilon_1$ and $\varepsilon_2$. Considering the TM polarization of electromagnetic wave, as shown in Fig. 6, it is possible to obtain two boundary conditions for the electric field $E^{(i)}$ and magnetic field $H^{(i)}$ $(i = 1, 2)$ as follows:

$$E_+^{(1)} \cos\vartheta + E_-^{(1)} \cos\vartheta = E_+^{(2)} \cos\phi \quad (24)$$
$$H_+^{(2)} - (H_+^{(1)} - H_-^{(1)}) = -\bar{\sigma} E_+^{(2)} \cos\phi \quad (25)$$

where +(-) index is the bottom- (top-) going waves according to Fig. 6, $\vartheta$ is the incident and reflection angle, and $\phi$ is the refraction angle. The $E$ and $H$ fields are related to each other in terms of the electromagnetic wave impedance, for each medium:

$$Z_i = \frac{E_i}{H_i} = \frac{377}{\sqrt{\varepsilon_i}}\ \Omega \quad (i = 1, 2) \quad (26)$$

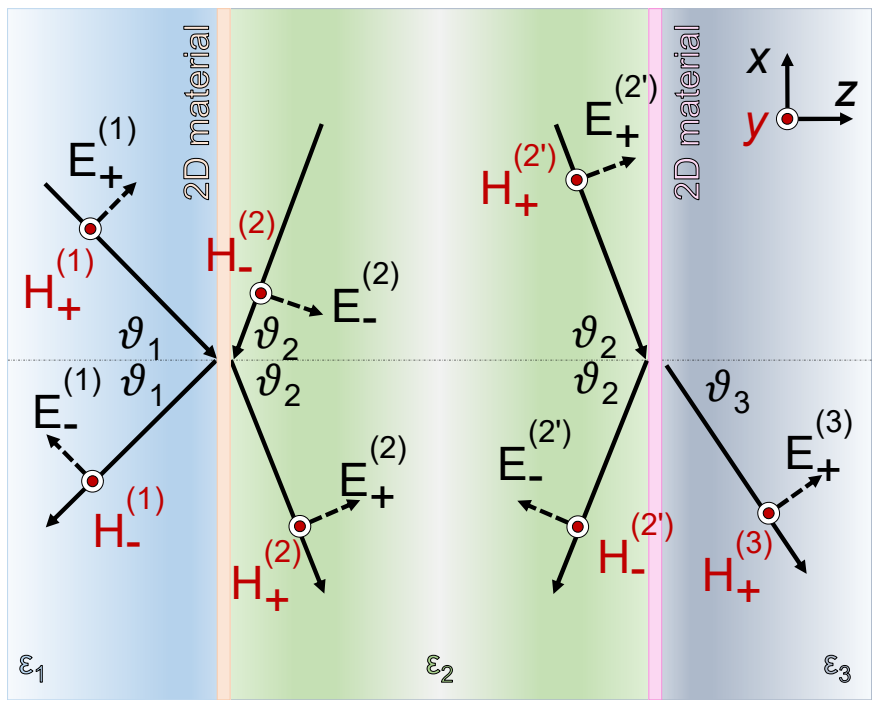

**FIG. 7:** The wave propagation through three media and two different 2D materials.

where the constant 377 Ω is the impedance of vacuum ($Z_0 = \sqrt{\mu_0/\varepsilon_0}$). Quantities $\phi$, $\vartheta$, and $Z_i$ are related by Snell's law ($Z_2 \sin\vartheta = Z_1 \sin\phi$). Solving Eqs. 24-26, reflection ($R$), transmission ($T$), and absorption ($A$) probabilities of the electromagnetic wave are obtained as follows:

$$R = \left| \frac{E_1^{(-)}}{E_1^{(+)}} \right|^2 = = \left| \frac{Z_2 \cos\phi - Z_1 \cos\vartheta - Z_1 Z_2 \bar{\sigma} \cos\vartheta \cos\phi}{Z_2 \cos\phi + Z_1 \cos\vartheta + Z_1 Z_2 \bar{\sigma} \cos\vartheta \cos\phi} \right|^2$$

$$T = \frac{\cos\phi}{\cos\vartheta} \frac{Z_1}{Z_2} \left| \frac{E_2^{(+)}}{E_1^{(+)}} \right|^2 = = \frac{4 Z_1 Z_2 \cos\vartheta \cos\phi}{|Z_2 \cos\phi + Z_1 \cos\vartheta + Z_1 Z_2 \bar{\sigma} \cos\vartheta \cos\phi|^2}$$

$$A = 1 - \mathrm{Re}\{R\} - \mathrm{Re}\{T\} = = \frac{4 Z_1 Z_2^2 \cos\vartheta \, |\cos\phi|^2 \, \mathrm{Re}\{\bar{\sigma}\}}{|Z_2 \cos\phi + Z_1 \cos\vartheta + Z_1 Z_2 \bar{\sigma} \cos\vartheta \cos\phi|^2} \tag{27}$$

where the values of $R$, $T$, and $A$ are real quantities and are denoted in terms of percentage ($0 - 100\%$).

*Multilayer configuration*

The configuration described in the previous section can be generalized to a multilayer configuration in which 2D and standard dielectric slubs alternate. This structure can be modeled usign the formalism of the equivalent chain of transmission lines [111], and then analyzed using the wave-amplitude transission matrix (WATM) [112]. The latter is very effective when, as in our case, we are mainly interested in an input-output description of the structure and has already been used in the present context [113–115]. Assuming $N$ 2D materials inside a multilayer dielectric media spaced from each other by a distance $z = d$, the WATM method [116] relates the electromagnetic fields at all interfaces in a multi-layered system. For simplicity, a system with two 2D material layers placed between three dielectric slubs (medium 1, 2 and 3 with dielectric constants $\varepsilon_1$, $\varepsilon_2$ and $\varepsilon_3$, respectively), can be assumed as a first approximation (Fig. 7). The incident electromagnetic wave with TM polarization comes from medium 1 with angle $\vartheta_1$ and it propagates through medium 2 with angle $\vartheta_2$ and medium 3 with angle $\vartheta_3$. By adopting the polarization TM of the incident electromagnetic wave, between two dielectric media ($0 < z < d$), the electromagnetic fields are related to each other through the boundary conditions obtained from Maxwell's equations. Therefore, it is possible to define the boundary conditions of the monolayer interface of the electric field $E^{(i)}$ and of the magnetic field $H^{(i)}$ ($i = 1, 2$) as follows:

$$E_x^{(1)} = E_x^{(2)} \tag{28}$$

$$H_y^{(2)} - H_y^{(1)} = -\bar{\sigma} E_x^{(2)} \tag{29}$$

Considering a thickness of the 2D material sufficiently small compared to the wavelength of the incident light ($d << \lambda$), the electric and magnetic fields in medium 1 can be described as follows:

$$E_x^{(1)}(z) = E_{x+}^{(1)}(z) + E_{x-}^{(1)}(z) = = E_{x_0+}^{(1)} e^{i\kappa_1 z} + E_{x_0-}^{(1)} e^{-i\kappa_1 z} \tag{30}$$

$$H_y^{(1)}(z) = \frac{\omega \varepsilon_0 \varepsilon_1}{\kappa_1} (E_{x_0+}^{(1)} e^{i\kappa_1 z} - E_{x_0-}^{(1)} e^{-i\kappa_1 z}) \tag{31}$$

and for medium 2:

$$E_x^{(2)}(z) = E_{x+}^{(2)}(z) + E_{x-}^{(2)}(z) = = E_{x_0+}^{(2)} e^{i\kappa_2 z} + E_{x_0-}^{(2)} e^{-i\kappa_2 z} \tag{32}$$

$$H_y^{(2)}(z) = \frac{\omega \varepsilon_0 \varepsilon_1}{\kappa_2} (E_{x_0+}^{(2)} e^{i\kappa_2 z} - E_{x_0-}^{(2)} e^{-i\kappa_2 z}) \tag{33}$$

where $E_x^{(i)}$ is the amplitude of the electric field along the $x$-axis inside medium $i$, $\varepsilon_i$ is the dielectric constant of the $i^{th}$ medium, $\omega$ is the angular frequency of the electromagnetic wave and $k_i$ is the wave vector on $z$-direction which is defined as:

$$k_i = \frac{2\pi}{\lambda} \sqrt{\varepsilon_i} \cos\vartheta_i \tag{34}$$

where $\lambda$ is the wavelength of the electromagnetic wave. The amplitude of electric field of the electromagnetic wave are $E_{x_0+}^{(i)}$ and $E_{x_0-}^{(i)}$ going to the $z$-direction. The reflected wave at the second interface with medium 3 inside the medium 2 is described by $E_{x_0-}^{(2)}$. The relationship between $H_y$ and $E_x$ comes from the following equation:

$$H_y^{(i)} = i\omega\varepsilon_0\varepsilon_i \int E_x^{(i)} dz \tag{35}$$

The corresponding angle of propagation of the electromagnetic wave inside each dielectric medium measured from the $z$-axis is $\vartheta_i$ ($i = 1, 2$). Quantities $\vartheta_i$ and $\varepsilon_i$ between two dielectric media are related by the Snell's law ($\varepsilon_1 \sin\vartheta_1 = \varepsilon_2 \sin\vartheta_2$). Using Eq. 28 through 33, by solving for $E_{x_0+}$ and $E_{x_0-}$ at the boundary explicitly, it is possible to write the following matrix:

$$\begin{bmatrix} E^{(1)}_{x_0+} \\ E^{(1)}_{x_0-} \end{bmatrix} = \left[M_1\right] \left[T_1\right] \left[M_2\right] \left[T_2\right] ... \left[T_{N-2}\right] \left[M_{N-1}\right] \begin{bmatrix} E^{(N)}_{x_0+} \\ 0 \end{bmatrix} \tag{36}$$

where:

$$M_i = \frac{1}{2} \begin{bmatrix} 1 + \Delta_i + G_i & 1 - \Delta_i + G_i \\ 1 - \Delta_i - G_i & 1 + \Delta_i - G_i \end{bmatrix} \tag{37}$$

$$T_i = \begin{bmatrix} e^{-i\kappa_{i+1}d} & 0 \\ 0 & e^{i\kappa_{i+1}d} \end{bmatrix} \tag{38}$$

Eq. 37 describes the electromagnetic waves in medium $i$ as a function of the electromagnetic waves in medium $i+1$ at the first boundary, where $\alpha_i$ and $\beta_i$ are:

$$G_i = \frac{\kappa_i \bar{\sigma}}{\omega \varepsilon_0 \varepsilon_i}, \quad \Delta_i = \frac{\kappa_i}{\kappa_{i+1}} \frac{\varepsilon_{i+1}}{\varepsilon_i} \tag{39}$$

Eq. 38 describes the propagation of the wave through the medium $i+1$, where $d$ is the length of the medium $i+1$.

The total reflection ($\rho_{total}$) and transmission ($\tau_{total}$) coefficients are:

$$\begin{bmatrix} E^{(1)}_{x_0+} \\ E^{(1)}_{x_0-} \end{bmatrix} = \begin{bmatrix} \xi & \zeta \\ \gamma & \delta \end{bmatrix} \begin{bmatrix} E^{(3)}_{x_0+} \\ 0 \end{bmatrix} \tag{40}$$

$$\rho_{total} = \frac{E^{(1)}_{x_0-}}{E^{(1)}_{x_0+}} = \frac{\gamma}{\xi}, \quad \text{and} \quad \tau_{total} = \frac{E^{(3)}_{x_0+}}{E^{(1)}_{x_0+}} = \frac{1}{\xi} \tag{41}$$

where $\xi$, $\zeta$, $\gamma$ and $\delta$ are the components of total matrix (Eq. 36), getting:

$$\begin{aligned} R &= \left|\rho_{total}\right|^2 \\ T &= \frac{\sqrt{\varepsilon_3}\cos\vartheta_1}{\sqrt{\varepsilon_1}\cos\vartheta_3} \left|\tau_{total}\right|^2 \\ A &= 1 - \mathrm{Re}\{R\} - \mathrm{Re}\{T\} \end{aligned} \tag{42}$$

where $R$, $T$, and $A$ are reflection, transmission, and absorption probabilities of the electromagnetic wave, respectively. In the case of the absence of a 2D material at the interface, $\alpha_i = 0$ can be set in the matching matrix of the corresponding interface.

With this theoretical background, it is possible to optimize the design of optical devices (*e.g.*, optical absorbers). For example, considering a device consisting of N 2D materials and respective media, the number and position of the absorption peaks can be tuned a priori, thus providing the theoretical basis for the study and design of the device and its application.

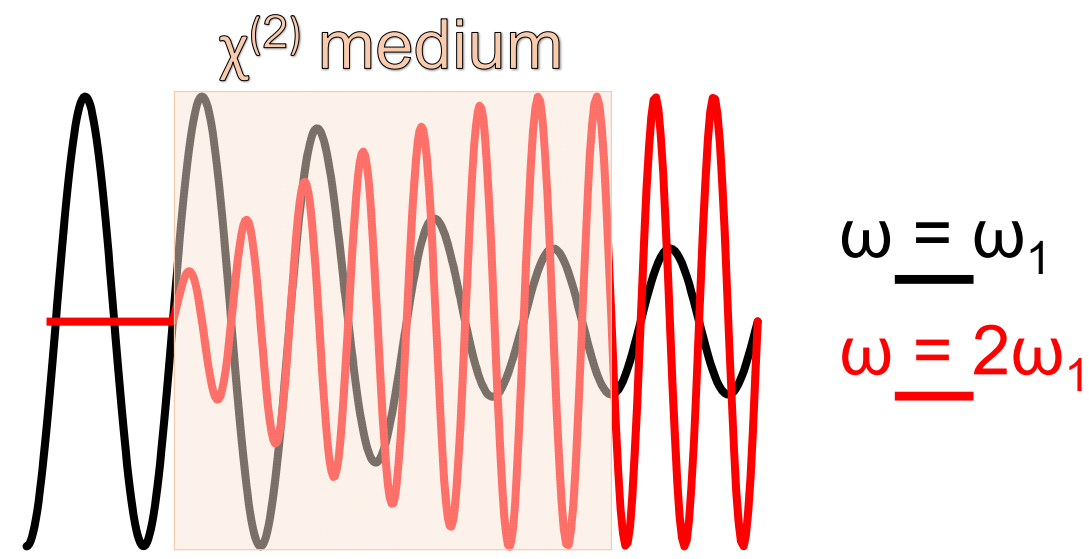

**FIG. 8:** Diagram of second-harmonic generation with perfect phase matching.

### Nonlinear optics in 2D materials

In the previous section, we discussed optical propagation in a 2D material. We have dealt with the problem of a linear dielectric medium impinged by an electromagnetic wave. In this context, a linear dielectric medium is characterized by a linear relation between the polarization density ($P$) and the electric field ($E$) [117]:

$$\boldsymbol{P} = \varepsilon_0 \boldsymbol{\chi} \boldsymbol{E} \tag{43}$$

where $\varepsilon_0$ is the permittivity of free space and $\chi$ is the electric susceptibility of the medium. The relation between $\boldsymbol{P}$ and $\boldsymbol{E}$ expressed by Eq. 43 describes the conventional linear optical effects, such as refraction and absorption and occurs when $\boldsymbol{E}$ is small, however, it becomes nonlinear when $\boldsymbol{E}$ acquires values comparable to interatomic electric fields, which are typically $\sim 10^5 - 10^8$ $Vm^{-1}$. Since externally applied optical electric fields are typically small in comparison with characteristic interatomic or crystalline fields, even when focused laser light is used, the nonlinearity is usually weak. A nonlinear anisotropic dielectric medium is characterized by a tensor relation which can be expanded in a Taylor series [117]:

$$P_i = \varepsilon_0 \left( \sum_j \chi_{ij} E_j + \sum_{jk} \chi^{(2)}_{ijk} E_j E_k + \right. \\ \left. + \sum_{jkl} \chi^{(3)}_{ijkl} E_j E_k E_l + ... \right) \tag{44}$$

where $i, j, k, l = 1, 2, 3$ and the coefficients $\chi_{ij}$, $\chi^{(2)}_{ijk}$ and $\chi^{(3)}_{ijkl}$ represent the tensor components of the $n^{th}$-order nonlinearity. These coefficients are characteristic constants of the medium and may be originated by microscopic or macroscopic phenomena. In particular, the $\boldsymbol{\chi}^{(2)}$ coefficient of the second term in Eq. 44, gives rise to second-order nonlinearity (three wave mixing), including second harmonic generation (SHG), sum and difference frequency generation (SFG, DFG), and optical parametric interaction (optical parametric amplification - OPA and optical parametric oscillation - OPO). Third order nonlinear effects, which usually arise from the susceptibility of third-order nonlinear optics ($\boldsymbol{\chi}^{(3)}$), include third-harmonic generation (THG), four-wave mixing (FWM), intensity-dependent refractive index change (optical Kerr effect and saturable absorption - SA) and two-photon excitation fluorescence (TPEF). The higher order coefficients ($\boldsymbol{\chi}^{(n)}$), represent the susceptibility of high-order nonlinear optics, namely the high-order multiphoton scattering/absorption/luminescence and the high-harmonic generation (HHG). Since the interaction intensity of nonlinear processes usually decreases with $n$ [118, 119], the effects of nonlinear higher-order effects (including HHG) have very low intensity, therefore typically $\boldsymbol{\chi}^{(n)} \sim 0$ for $n > 3$. In the case of bulk crystals, the generation of nonlinear effects occurs when the phase-matching condition is satisfied, in order to maximize the intensity of the nonlinear optical signal. According to quantum mechanics, the phase-matching condition is satisfied when the photon's momentum and energy are simultaneously conserved before and after the nonlinear process [119]. For this reason, the path of the incident light and the orientation of the crystal must be carefully designed in order to optimize the non-linear effects [120]. However, in the case of a medium with reduced thickness, comparable to the sub-wavelength range, and shorter than the coherence length, the phase matching condition is easily achieved. This allows obtaining strong nonlinear effects [120]. Furthermore, the crystalline structure of the material strongly influences nonlinear processes. In fact, thanks to the high values of the tensors $\boldsymbol{\chi}^{(2)}$ and $\boldsymbol{\chi}^{(3)}$, some materials show an anisotropic nonlinear response [121] even if the linear optical response is isotropic. This allows characterizing the orientation of the crystal with very high sensitivity [118, 119].

Since the susceptibility $\boldsymbol{\chi}^{(2)}$ is a third-rank tensor of even parity under inversion, $\boldsymbol{\chi}^{(2)}$ vanishes for media with inversion symmetry. Namely, in centrosymmetric media, which have inversion symmetry, the properties of the medium are not altered by the transformation $\boldsymbol{r} \to -\boldsymbol{r}$ [117]; the $\boldsymbol{P} - \boldsymbol{E}$ function must have odd symmetry so that the reversal of $E$ results in the reversal of $P$ without any other change:

$$\begin{aligned} \boldsymbol{P} = \varepsilon_0 \left( \boldsymbol{\chi}\boldsymbol{E} + \boldsymbol{\chi}^{(2)}\boldsymbol{E}^2 + \boldsymbol{\chi}^{(3)}\boldsymbol{E}^3 \right) = \\ = \varepsilon_0 \left( \boldsymbol{\chi}(-\boldsymbol{E}) + \boldsymbol{\chi}^{(2)}(-\boldsymbol{E})^2 + \boldsymbol{\chi}^{(3)}(-\boldsymbol{E})^3 \right) = \\ = -\boldsymbol{P} \end{aligned} \tag{45}$$

In order to verify the latter, the second-order nonlinear susceptibility coefficient must be zero ($\boldsymbol{\chi}^{(2)} = 0$). In contrast, odd-order nonlinearity (*i.e.* $\boldsymbol{\chi}^{(3)}$) is allowed in any material, regardless of whether the material is centrosymmetric, such as THG and FWM [119].

Given a plane wave of amplitude $E(\omega)$ traveling in a nonlinear medium (in this case a non-centrosymmetric 2D material) in the direction of its $q$ vector (Fig. 8), a polarization is generated at the second-harmonic frequency as follows:

$$P(2\omega) = \varepsilon_0 \chi^{(2)} E^2(\omega) = 2\varepsilon_0 d_{eff} E^2(\omega) \tag{46}$$

where $d_{eff}$ is the effective nonlinear optical coefficient which depends on $\chi^{(2)}$.

If the slowly varying envelope approximation is applied [122]:

$$\left| \frac{\partial^2 E(z,\omega)}{\partial z^2} \right| << k \left| \frac{\partial E(z,\omega)}{\partial z} \right| \tag{47}$$

and assuming negligible losses, the wave equation at $2\omega$ is:

$$\frac{\partial E(z,2\omega)}{\partial z} = -\frac{i\omega}{n_{2\omega} c} d_{eff} E^2(z,\omega) e^{i\Delta q z} \tag{48}$$

where $\Delta q = q(2\omega) - 2q(\omega)$. Considering the low conversion efficiency assumption ($E(2\omega) << E(\omega)$), which is a valid condition when the conversion to the second harmonic is not significant, the amplitude $E(\omega)$ remains constant over the thickness of the nonlinear medium, $t$. Hence, by using the boundary condition $E(2\omega, z = 0) = 0$, we get:

$$\begin{aligned} E(2\omega, z = t) = -\frac{i\omega}{n_{2\omega} c} d_{eff} E^2(z,\omega) \int_0^t e^{i\Delta q z}\, dz = \\ = -\frac{i\omega}{n_{2\omega} c} d_{eff} E^2(\omega) t \frac{\sin\left(\frac{1}{2}\Delta q t\right)}{\frac{1}{2}\Delta q t} e^{\frac{i}{2}\Delta q t} \end{aligned} \tag{49}$$

In terms of the optical intensity:

$$I(2\omega, t) = |E(2\omega, t)|^2 = \frac{2\omega^2 d_{eff}^2 t^2}{n_{2\omega} n_\omega^2 c^3 \varepsilon_0} \left( \frac{\sin\left(\frac{1}{2}\Delta q t\right)}{\frac{1}{2}\Delta q t} \right)^2 I^2(\omega) \tag{50}$$

The intensity $I(2\omega)$ is maximized for the phase-matched condition $\Delta q = 0$. The Eq. 50 shows that $I(2\omega, t) = 0$ when $d_{eff} \propto \chi^{(2)} = 0$ for media presenting inversion symmetry. For example, the natural vertical stacking of the most commonly used transition metal dichalcogenide semiconductors - TMDS (2H polytype), is constituted by monoatomic layers rotated by 180° with respect to their next neighbors, forming the so-called AB stack [123]. As a direct consequence, the inversion symmetry is present for even layers ($\chi^{(2)} = 0$), preventing the observation of any second-order nonlinear process [124]. However, TMDS can be reduced to nanometric thickness ($t \sim 1$ $nm$) due to the weak van der Waals interlayer forces, showing a non-centrosymmetric structure ($\chi^{(2)} \neq 0$) in their monolayer form. In this case, the medium is usually presented without any phase factor ($\Delta q = 0$) due to the atomically thin nature [125, 126], hence:

$$E(2\omega) = -\frac{i\omega}{n_{2\omega} c} d_{eff} E^2(\omega) t \tag{51}$$

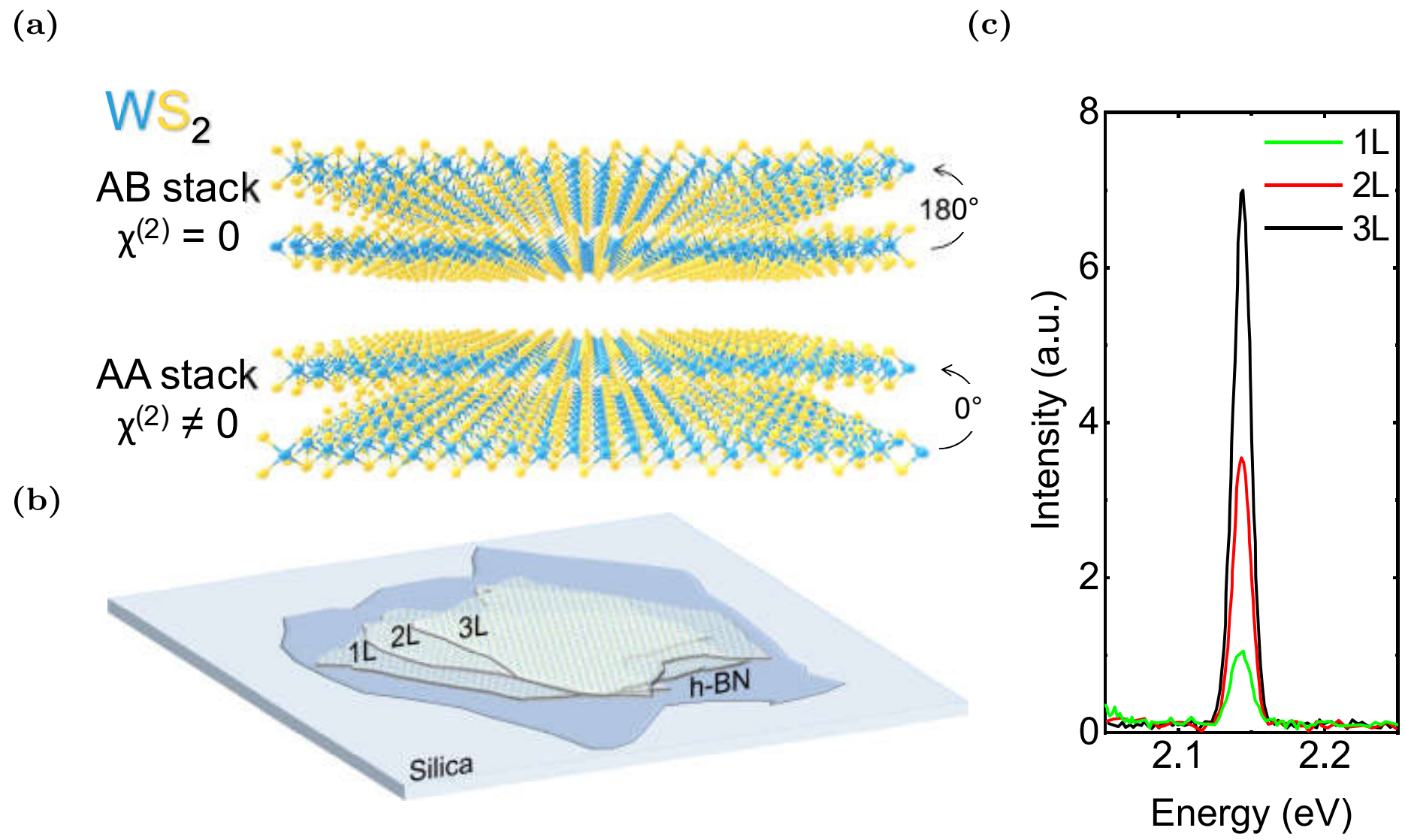

**FIG. 9:** a) Schematics of natural centrosymmetric AB structure with an interlayer twist angle of 180° ($\chi^{(2)} = 0$) and artificial non-centrosymmetric AA configuration with 0° twist angle ($\chi^{(2)} \neq 0$). b) Sketch of the $WS_2$ AA stack sample. c) Photoluminescence spectra from as a function of the number of layers. Data from ref. [122].

However, the crystal structure is very important for symmetry reasons and it is possible to find bulk crystals of TMDS possessing a $\chi^{(2)} \neq 0$. For example, in ref. [126] the authors probe the SHG generated from the noncentrosymmetric $3R$ crystal phase of $MoS_2$. Instead Trovatello et. al., in ref. [122] show that artificial crystals with vertically stacked $WS_2$ monolayers, with interlayer $\sim 0°$ crystal angle alignment, forming the so-called AA stack [123], preserve the broken inversion symmetry ($\chi^{(2)} \neq 0$), as shown in Fig. 9a. In both cases, if no strong interactions are assumed between the layers and each layer is treated as optically independent, it is possible to model the individual electric-field from any layer as:

$$E(N, 2\omega) \propto e^{i\Delta q(N-1)t} e^{-\alpha(N-1)t/2} \tag{52}$$

where $N$ is the number of the layers and $(N-1)$ results due to the indexing of the crystal, *i.e.*, the top-most layer is 1, so the subsequent intensity is normalized to 1, and $t$ is the crystal thickness. To account for the re-absorption of the second harmonic light, it is introduced an exponential loss factor, $\alpha$. This is the attenuation factor at the second harmonic energy extracted from the single-layer linear absorption spectrum ($i.e.$, $\gamma = 1 - I_{abs} = e^{-\alpha t}$). It is possible to write the total SH light from $N$ layers as:

$$I(N, 2\omega) \propto \left| \sum_{M=1}^{N} e^{i\Delta q(M-1)t} e^{-\alpha(M-1)t/2} \right|^2 \tag{53}$$

From Eq. 53, it is clear that the intensity scales quadratically with $N$ in the case of a non-centrosymmetric crystal of $N$ layers. This has been observed experimentally, performing a proof-of-principle demonstration. As shown in Fig. 9b, three layers of manually AA stacked $WS_2$ sitting on top of a h−BN flake, have been identified by photoluminescence characterization, demonstrating that artificial stacking of AA aligned $WS_2$ monolayers provides a route for quadratic scaling of the efficiency with layers' number (Fig. 9c).

## CONCLUSION

2D materials offer compelling perspectives for implementing devices with new capabilities over those based on conventional materials. In non-interacting 2D material layers, the transfer matrix method can be used to obtain their reflectance, transmittance, and absorbance spectra. In this way, photonic band structures for periodical 2D material layers can be studied in a simple way. Another key feature of 2D non-linear materials is their non-centrosymmetry, which can be easily realized in 2D materials. For example, $WS_2$ becomes non-centrosymmetric when its bulk counterpart is thinned into a monolayer. These properties demonstrate that these effects greatly improve performance in 2D material-based devices. Hence, it is of great importance to pursue appropriate 2D crystals without thickness limitation, which can provide more opportunities to obtain high-performance optoelectronic devices. All this research paves the way for optical devices ranging from sensors to modulators to isolators, serving as a versatile tool to study optical properties in 2D materials.